\documentclass[11pt,a4paper]{article}

\pdfoutput=1

\usepackage{jheppub}
\usepackage{amsfonts}
\usepackage{amssymb}
\usepackage{graphicx}
\usepackage{tikz}
\usepackage{url}
\usepackage{hyperref}
\hypersetup{hypertexnames=false}

\newcommand{\be}{\begin{equation}}
\newcommand{\ee}{\end{equation}}
\newcommand{\bea}{\begin{eqnarray}}
\newcommand{\eea}{\end{eqnarray}}

\title{{\color{blue} Constraining Axion-Like-Particles with Hard X-ray Emission from Magnetars}}

\author{Jean-Fran\c{c}ois Fortin$^1$ and Kuver Sinha$^2$}
\affiliation{
$^1$D\'epartement de Physique, de G\'enie Physique et d'Optique, Universit\'e Laval,\\\phantom{$^1$}Qu\'ebec, QC G1V 0A6, Canada\\
$^2$Department of Physics and Astronomy, University of Oklahoma, Norman, OK 73019, USA}

\abstract{Axion-like particles (ALPs) produced in the core of a magnetar will convert to photons in the magnetosphere, leading to possible signatures in the hard X-ray band.  We perform a detailed calculation of the ALP-to-photon conversion probability in the magnetosphere, recasting the coupled differential equations that describe ALP-photon propagation into a form that is efficient for large scale numerical scans.  We show the dependence of the conversion probability on the ALP energy, mass, ALP-photon coupling, magnetar radius, surface magnetic field, and the angle between the magnetic field and direction of propagation.  Along the way, we develop an analytic formalism to perform similar calculations in more general $n$-state oscillation systems.  Assuming ALP emission rates from the core that are just subdominant to neutrino emission, we calculate the resulting constraints on the ALP mass versus ALP-photon coupling space, taking SGR 1806-20 as an example.  In particular, we take benchmark values for the magnetar radius and core temperature, and constrain the ALP parameter space by the requirement that the luminosity from ALP-to-photon conversion should not exceed the total observed luminosity from the magnetar.  The resulting constraints are competitive with constraints from helioscope experiments in the relevant part of ALP parameter space.}

\begin{document}
\maketitle


\section{Introduction}

Magnetars constitute an interesting subclass of neutron stars characterized by extremely strong magnetic fields.  Assuming that their rapid rotational spin down is caused by magnetic dipole torques, X-ray timing properties of magnetars yield field values that generally exceed the quantum critical value $B_c=m_e^2/e=4.414\times10^{13}\,\text{G}$.  For recent reviews of magnetars, we refer to \cite{Turolla:2015mwa,Beloborodov:2016mmx,Kaspi:2017fwg}.

The extreme magnetic field of magnetars can potentially be exploited in the search for new physics beyond the Standard Model.  Clearly, the most fertile area to investigate is the physics of axions, originally introduced as a solution to the strong CP problem \cite{Weinberg:1977ma,Kim:1979if,Dine:1981rt}.  Our interest in this paper will be in the broader class of axion-like particles (ALPs) which arise generically in compactifications in string theory, and have been applied to phenomena ranging from inflation to baryogenesis \cite{Arvanitaki:2009fg}.  In the presence of an external magnetic field, ALPs can convert into photons, and vice versa, via the Primakoff process.  The relevant coupling is given by the first term in
\be\label{gagamma}
\mathcal{L}\supset-\frac{g}{4}aF_{\mu\nu}\tilde{F}^{\mu\nu}+g_{aN}(\partial_{\mu}a)\bar{N}\gamma^{\mu}\gamma_5N,
\ee
where $a$ denotes the ALP and the coupling $g\equiv g_{a\gamma}$ has mass dimension $-1$.  This process lies at the heart of many ALP searches and constraints.  The second term in \eqref{gagamma}, which is model-dependent, describes the coupling between the ALP $a$ and nucleons $N$, and leads (among other processes) to ALP production in the core of neutron stars.  For reviews, we refer to \cite{Marsh:2015xka,Graham:2015ouw}.  

Photons (ALPs) emitted by a magnetar will convert to ALPs (photons) via Eq.~\eqref{gagamma}.  It is therefore interesting to investigate whether observations of magnetars can constrain the ALP mass and coupling strength $g$.  This avenue has been pursued by several authors recently \cite{Lai:2006af,Chelouche:2008ta,Jimenez:2011pg,Perna:2012wn}.\footnote{Magnetars have also been studied recently by particle physicists in other contexts, for example to constrain milli-magnetically charged particles \cite{Hook:2017vyc} and millicharged fermions \cite{Korwar:2017dio}.}  We briefly summarize the findings of \cite{Lai:2006af} and \cite{Perna:2012wn}, which are germane to the present work.

The authors of \cite{Lai:2006af} and \cite{Perna:2012wn} studied the surface emission of photons from neutron stars and their subsequent conversion into ALPs in the magnetosphere.  Modifications of the spectral shape, light curves, and polarization signals that could potentially be identified observationally were investigated after taking into account effects of gravitational redshift and light deflection.  For a star that emits from its entire surface, these features include deviations in the apparent radius and emission area from limits set by the neutron star equations of state.  Moreover, interesting polarization signals can also arise -- for example, an emission region that is observed phase-on will exhibit an inversion of the plane of polarization compared to the case where photon-axion conversion is absent \cite{Perna:2012wn}.  The features studied in these papers concentrate on the soft X-ray emission from neutron stars.

We note that the general strategy of these papers, and our current work, is different from usual cooling arguments that are used to place limits on ALPs from white dwarfs \cite{Dreiner:2013tja} or supernovas \cite{Chang:2018rso}.  While attempts have been made to use cooling simulations of neutron stars to constrain ALPs \cite{Sedrakian:2015krq}, our work relies specifically on the Primakoff process in the magnetosphere.

The purpose of our paper is to understand the following question: can ALPs produced inside the neutron star lead to observational signatures via \textit{ALP-to-photon} conversion in the magnetosphere?  There are two aspects to this question, and we outline our approach to them as follows:

\begin{itemize}
\item[$(i)$]The first aspect concerns the probability of ALP-to-photon conversion in the magnetosphere, and its dependence on the ALP parameters and the properties of the magnetar.  We have performed a careful analysis of the coupled differential equations describing the ALP-photon system as it propagates through the magnetosphere.  We have recast the equations into a form that is efficient for large scale numerical scans, and checked that there is a gain of almost two orders of magnitude in computation time.  Along the way, we develop an analytic formalism to perform similar calculations in more general $n$-state oscillation systems.  We provide detailed scans of the conversion probability as a function of the ALP energy $\omega$ in the $1-200\,\text{keV}$ range, mass $m_a$, coupling $g$, surface magnetic field strength $B_0$ of the magnetar, magnetar radius $r_0$, and angle between the magnetic field and the direction of propagation $\theta$.
\item[$(ii)$]The second aspect concerns the emission rate of ALPs from the magnetar.  We assume that ALPs are emitted uniformly from the core of the neutron star and escape into the magnetosphere.  Furthermore, we assume that the emission rate is just subdominant to the neutrino emission rate for a given core temperature.  This conservatively leaves standard cooling mechanisms of neutron stars unaltered, while also giving the best case scenario for signals of ALP-to-photon conversion in the magnetosphere.  We take nucleon-nucleon bremsstrahlung $N+N\to N+N+a$ as the main production mechanism of ALPs, while noting that various other production mechanisms have been considered in the literature \cite{Iwamoto:1984ir}.  We reserve a more detailed study of this question and especially its relation to ALP-photon conversion for a future publication.
\end{itemize}

Having calculated the ALP-to-photon conversion probability and assumed the ALP emission rate from the core, we study possible signatures of the resulting photon flux.  The ALP spectrum from nucleon-nucleon bremsstrahlung is a broad distribution peaked at energies $\omega\sim3.3T$ where $T$ is the magnetar core temperature, assuming degenerate nuclear matter in the interior \cite{Raffelt:1996wa}.  With core temperatures $T$ in the range $(0.6-3.0)\times10^9\,\text{K}\sim50-250\,\text{keV}$, the hard X-ray band $100-900\,\text{keV}$ should be the natural focus to investigate unique features of ALP-to-photon conversion.

Since ALPs are \textit{steadily} produced by the core and converted to photons in the magnetosphere, we will mainly be interested in quiescent magnetar emission.\footnote{In particular, we will have nothing to say about the hard X-ray outbursts or giant supersecond flares shown by some SGRs and AXPs.}  Both Soft-Gamma Repeaters (SGRs) and Anomalous X-ray Pulsars (AXPs) exhibit quiescent soft X-ray emission below 10 keV.  This mostly thermal emission with luminosity $L\sim10^{33}-4\times10^{35}\,\text{erg}\cdot\text{s}^{-1}$ is presumably powered by the internal magnetic energy \cite{Vigano:2013lea}.  The spectra of several AXPs and SGRs also show hard non-thermal tails extending up to $150-200\,\text{keV}$ \cite{Kuiper:2004ya,Hartog:2008tp,Gotz:2006cx,Enoto:2010ix} with luminosities similar to those observed in the soft X-ray band.  While the soft X-ray emission is thermal and originates from the hot surface of the magnetar, the origin of the hard X-ray spectrum is an area of active study.  We refer to \cite{Wadiasingh:2017rcq,Wadiasingh:2017nrr,Beloborodov2012} for more details. 

ALPs produced from the core and converting to photons in the magnetosphere can contribute significantly to the spectral peak above $100\,\text{keV}$ for optimal selections of ALP parameters.  The spectral peak given by $\omega\sim3.3T$ can be anywhere up to almost $1\,\text{MeV}$ and for optimal selections of $m_a$ and $g$, one can match the observed luminosities.  Of course, a detailed analysis would be required before one can claim that this is the \textit{dominant} contribution or mechanism underlying hard X-ray emissions from magnetars.

Our far more conservative approach in this paper will be to constrain ALP parameters using the observed hard X-ray emission.  In other words, we will demand that the luminosity coming from ALP-to-photon conversion be bounded by the observed luminosity in the range $1-200\,\text{keV}$, while remaining agnostic about the physical processes that give rise to this emission.  We will take the benchmark value of the magnetic field of SGR 1806-20 \cite{Palmer:2005mi}, although other candidates from the McGill Magnetar Catalog \cite{Olausen:2013bpa} can also be considered.  For the magnetar radius, we assume a typical value of $r_0=10\,\text{km}$, while for the core temperature, we assume a range of values between $T_9=0.6$ to $T_9=3.0$ where $T_9=T/(10^9\,\text{K})$.  We then compute the predicted photon flux from ALP-to-photon conversion in the entire energy range from $\omega=1-200\,\text{keV}$ and compare this against the observed average luminosity of SGR 1806-20 in this band.  This places limits on the parameter space of ALP masses $m_a$ and ALP-photon coupling $g$.  The limits we obtain are competitive with constraints from helioscope experiments in the relevant part of ALP parameter space for high enough core temperatures.  We reserve a more detailed spectral analysis, as well as possible polarization signals in the hard X-ray regime, for a future publication.

The rest of the paper is structured as follows.  In Section \ref{apconvprob}, we provide the calculation of the ALP-to-photon conversion probability from a general formalism for $n$-state oscillation problems.  In Sections \ref{benchmark} and \ref{SSGR}, we apply our results to SGR 1806-20, and present the constraints on the ALP parameter space.  The details of ALP production in the core are relegated to an appendix.


\section{ALP-to-Photon Conversion Probability}\label{apconvprob}

This section develops a general formalism for oscillation in the weak-dispersion limit.  We first give an analytic treatment of $n$-state oscillation problems using conservation of probability.  We then apply this formalism to ALP-photon oscillations in the magnetosphere of magnetars.  Our main goal here is to recast the standard evolution equations in the weak-dispersion limit of \cite{Lai:2006af} into a formalism that is more amenable for numerical applications, such as large scans over parameter space.  We have checked that our formalism leads to integration times that are faster by around two orders of magnitude.

\subsection{General Formalism}

For a general $n$-state oscillation problem in the weak-dispersion limit \cite{Raffelt:1987im}, the coupled system of differential equations for the linearized wave equations is of the form
\be
i\frac{da_i(x)}{dx}=\sum_{j=1}^nA_{ij}(x)a_j(x),
\ee
where $a_i(x)$ are the oscillating fields and $A_{ij}(x)$ is some matrix dictated by the oscillation problem at hand.  Conservation of probability $\frac{d}{dx}\sum_{i=1}^n|a_i(x)|^2=0$ implies that $A_{ji}(x)=A_{ij}(x)$.\footnote{More generally, $A_{ji}^*(x)=A_{ij}(x)$ also leads to conservation of probability.  The remainder of the analysis will be performed for a real-symmetric $A(x)$ since $A(x)$ is real-symmetric for ALP-photon conversion \cite{Lai:2006af}.  The generalization to hermitian $A(x)$ is straightforward and is left to the reader.}  Using generalized spherical coordinates, one possible choice for the form of the solutions is
\be
a_i(x)=\left\{\prod_{j=i}^{n-1}\sin[\chi_j(x)]\right\}\cos[\chi_{i-1}(x)]e^{-i\phi_i(x)},\
\ee
with $\chi_0(x)=0$.  This form satisfies conservation of probability such that the total probability is properly normalized, \textit{i.e.} $\sum_{i=1}^n|a_i(x)|^2=1$.

Differentiating with respect to $x$ gives
\be
i\frac{da_i(x)}{dx}=i\left\{\sum_{j=i}^{n-1}\cot[\chi_j(x)]\frac{d\chi_j(x)}{dx}-\tan[\chi_{i-1}(x)]\frac{d\chi_{i-1}(x)}{dx}-i\frac{d\phi_i(x)}{dx}\right\}a_i(x),
\ee
and thus the coupled system of differential equations becomes
\be
i\sum_{j=i}^{n-1}\cot[\chi_j(x)]\frac{d\chi_j(x)}{dx}-i\tan[\chi_{i-1}(x)]\frac{d\chi_{i-1}(x)}{dx}+\frac{d\phi_i(x)}{dx}=\frac{1}{a_i(x)}\sum_{j=1}^nA_{ij}(x)a_j(x).
\ee
The right-hand side can be expressed in terms of the new variables as
\be
\frac{1}{a_i(x)}\sum_{j=1}^nA_{ij}(x)a_j(x)=\sum_{j=1}^nA_{ij}(x)\frac{\left\{\prod_{k=j}^{n-1}\sin[\chi_k(x)]\right\}\cos[\chi_{j-1}(x)]}{\left\{\prod_{k=i}^{n-1}\sin[\chi_k(x)]\right\}\cos[\chi_{i-1}(x)]}e^{-i[\phi_j(x)-\phi_i(x)]}.\nonumber
\ee
Hence, gathering real and imaginary parts, the coupled system of differential equations simplifies further to
\bea
\frac{d\chi_{i-1}(x)}{dx}&=&\sum_{j=1}^nA_{ij}(x)S_{ij}(x)\nonumber\\
&\phantom{=}&\hspace{1cm}+\cot[\chi_{i-1}(x)]\sum_{j=i+1}^n\left\{\sum_{\ell=1}^nA_{j\ell}(x)S_{j\ell}(x)\right\}\cot[\chi_{j-1}(x)]\left\{\prod_{k=i}^{j-2}\csc^2[\chi_k(x)]\right\},\nonumber\\
\frac{d\phi_i(x)}{dx}&=&\sum_{j=1}^nA_{ij}(x)C_{ij}(x).\nonumber
\eea
where
\bea
S_{ij}(x)&=&\frac{\left\{\prod_{k=j}^{n-1}\sin[\chi_k(x)]\right\}\cos[\chi_{j-1}(x)]}{\left\{\prod_{k=i}^{n-1}\sin[\chi_k(x)]\right\}\sin[\chi_{i-1}(x)]}\sin[\phi_j(x)-\phi_i(x)],\nonumber\\
C_{ij}(x)&=&\frac{\left\{\prod_{k=j}^{n-1}\sin[\chi_k(x)]\right\}\cos[\chi_{j-1}(x)]}{\left\{\prod_{k=i}^{n-1}\sin[\chi_k(x)]\right\}\cos[\chi_{i-1}(x)]}\cos[\phi_j(x)-\phi_i(x)].
\eea
It is interesting to note that the evolution equations are not functions of the $n$ individual phases $\phi_i(x)$, but only of the $n-1$ phase differences $\Delta\phi_{ij}(x)=\phi_i(x)-\phi_j(x)$.  Including the $n-1$ angles $\chi_i(x)$, this leads to $2(n-1)$ real differential equations instead of $2n$ real differential equations, as expected from conservation of probability.

Before proceeding, it is also important to mention that the general formalism introduced here can be used for any oscillation problem.  Hence, it might be useful in the study of neutrino oscillations.

\subsection{ALP-Photon Oscillations}

We now focus on ALP-photon oscillations in a magnetic field \cite{Lai:2006af}, as for example the magnetic field of a magnetar.  As long as the magnetic field space variations occur on larger distances than the photon (and ALP) wavelength and the magnetic field is not too large, the ALP-to-photon conversion probability can be calculated in the weak-dispersion limit and the general formalism developed above can be used directly.  Moreover, since the vacuum resonance and the ALP-photon resonance are well separated in the magnetized plasma of a magnetar in the ALP parameter space of interest here, the perpendicular photon electric field mode decouples from the evolution equations and can be forgotten altogether \cite{Lai:2006af}.

Hence, the coupled system of differential equations describing ALP-photon oscillations is given by
\be\label{EqnDiffDim}
i\frac{d}{dr}\left(\begin{array}{c}a\\E_\parallel\end{array}\right)=\left(\begin{array}{cc}\omega+\Delta_a&\Delta_M\\\Delta_M&\omega+\Delta_\parallel\end{array}\right)\left(\begin{array}{c}a\\E_\parallel\end{array}\right),
\ee
where $\Delta_\parallel$ and $\Delta_M$ are functions of $r$,
\be
\Delta_a=-\frac{m_a^2}{2\omega},\qquad\qquad\Delta_\parallel=\frac{1}{2}q\omega\sin^2\theta,\qquad\qquad\Delta_M=\frac{1}{2}gB\sin\theta.\nonumber
\ee
Here $a(r)$ and $E_\parallel(r)$ are the ALP and parallel photon electric fields respectively, $r$ is the distance from the center of the magnetar, $\omega$ is the energy of the ALP and photon electric fields, $m_a$ is the ALP mass, $g$ is the ALP-photon coupling constant, $\theta$ is the angle between the direction of propagation of the ALP-photon field and the magnetic field, and $q$ is a dimensionless function of the magnetic field $B$ given by \cite{Lai:2006af,Raffelt:1987im}
\be
q=\frac{7\alpha}{45\pi}b^2\hat{q},\qquad\qquad\hat{q}=\frac{1+1.2b}{1+1.33b+0.56b^2},\nonumber
\ee
with $b=B/B_c$ where $B_c=m_e^2/e=4.414\times10^{13}\,\text{G}$ is the critical QED field strength.  Here $e=\sqrt{4\pi\alpha}$ where the fine structure constant is approximatively $\alpha\approx1/137$.

With the help of the dimensionless variable $x=r/r_0$ where $r_0$ is the magnetar radius, the coupled system of differential equations \eqref{EqnDiffDim} becomes
\be\label{EqnDiff}
i\frac{d}{dx}\left(\begin{array}{c}a\\E_\parallel\end{array}\right)=\left(\begin{array}{cc}\omega r_0+\Delta_ar_0&\Delta_Mr_0\\\Delta_Mr_0&\omega r_0+\Delta_\parallel r_0\end{array}\right)\left(\begin{array}{c}a\\E_\parallel\end{array}\right)=\left(\begin{array}{cc}A(x)&D(x)\\D(x)&B(x)\end{array}\right)\left(\begin{array}{c}a\\E_\parallel\end{array}\right).
\ee
The relevant matrix for ALP-photon oscillation is therefore real-symmetric with $A(x)\equiv A$.  Hence the general formalism developed above can be used directly.

However, for further convenience, we choose solutions of the form
\be\label{EqnSoln}
a(x)=\cos[\chi(x)]e^{-i\phi_a(x)},\qquad\qquad E_\parallel(x)=i\sin[\chi(x)]e^{-i\phi_E(x)},
\ee
with $\chi(x)$, $\phi_a(x)$ and $\phi_E(x)$ real functions.  The extra phase in \eqref{EqnSoln} will simplify the initial conditions for pure initial states.  Thus the coupled system of differential equations \eqref{EqnDiff} simplifies to
\bea
\frac{d\chi(x)}{dx}+i\cot[\chi(x)]\left[\frac{d\phi_a(x)}{dx}-A(x)\right]&=&-D(x)e^{i[\phi_a(x)-\phi_E(x)]},\nonumber\\
\frac{d\chi(x)}{dx}-i\tan[\chi(x)]\left[\frac{d\phi_E(x)}{dx}-B(x)\right]&=&-D(x)e^{-i[\phi_a(x)-\phi_E(x)]}.\nonumber
\eea
Gathering real and imaginary parts, the coupled system of differential equations simplifies further to
\bea
\frac{d\chi(x)}{dx}&=&-D(x)\cos[\phi_a(x)-\phi_E(x)],\nonumber\\
\frac{d\phi_a(x)}{dx}&=&A(x)-D(x)\tan[\chi(x)]\sin[\phi_a(x)-\phi_E(x)],\nonumber\\
\frac{d\phi_E(x)}{dx}&=&B(x)-D(x)\cot[\chi(x)]\sin[\phi_a(x)-\phi_E(x)].\nonumber
\eea
Again, the fact that both real parts lead to the same differential equation confirms the form \eqref{EqnSoln} as expected from arguments about conservation of probability.

Since the focus is on the conversion probability and only the relative phase $\Delta\phi(x)=\phi_a(x)-\phi_E(x)$ appears in the equations above, one gets to
\be\label{EqnEvol}
\begin{aligned}
\frac{d\chi(x)}{dx}&=-D(x)\cos[\Delta\phi(x)],\\
\frac{d\Delta\phi(x)}{dx}&=A(x)-B(x)+2D(x)\cot[2\chi(x)]\sin[\Delta\phi(x)],
\end{aligned}
\ee
where $\chi(1)$ determines the initial state at the surface of the magnetar.  To avoid singularities for $\chi(1)=n\pi/2$ with $n\in\mathbb{Z}$, \textit{i.e.} for pure initial states, the initial condition for $\Delta\phi(1)$ must satisfy $\Delta\phi(1)=m\pi$ with $m\in\mathbb{Z}$.  It is therefore possible to set $\Delta\phi(1)=0$ for a pure ALP initial state\footnote{This is possible thanks to the choice of solutions \eqref{EqnSoln}.} and the ALP-photon conversion probability is simply $P_{a\to\gamma}(x)=\sin^2[\chi(x)]$.

\subsection{Implications for Conversion Probability}

Unfortunately, \eqref{EqnEvol} cannot be solved analytically for generic cases.  Nevertheless, exact solutions exist for some specific cases.  For example, in the no-mixing case, $D(x)=0$ and solutions to \eqref{EqnEvol} are
\be\label{EqnSolnC0}
\chi(x)=\chi(1),\qquad\qquad\Delta\phi(x)=\Delta\phi(1)+\int_1^xdx'\,[A(x')-B(x')],
\ee
showing explicitly that the initial amplitudes do not change as the state propagates.  Moreover, there exist exact solutions to the evolution equations \eqref{EqnEvol} in the case $A(x)=B(x)$ for $\Delta\phi(1)=0$.  Indeed, in this specific case it is easy to see that
\be\label{EqnSolnAB}
\chi(x)=\chi(1)-\int_1^xdx'\,D(x'),\qquad\qquad\Delta\phi(x)=0,
\ee
are solutions to \eqref{EqnEvol}.  Therefore, when $A(x)=B(x)$ the phase difference stays constant while $\chi(x)-\chi(1)$ varies appropriately.

One can also write a general solution for $\chi(x)$ in terms of the functions $D(x)$ and $\Delta\phi(x)$ as
\be\label{Eqnchi}
\chi(x)=\chi(1)-\int_1^xdx'\,D(x')\cos[\Delta\phi(x')].
\ee
The solution \eqref{Eqnchi} leads to an upper bound given by
\be\label{EqnBound}
|\chi(x)-\chi(1)|=\left|\int_1^xdx'\,D(x')\cos[\Delta\phi(x')]\right|\leq\int_1^xdx'\,|D(x')||\cos[\Delta\phi(x')]|\leq\int_1^xdx'\,|D(x')|.
\ee
The upper bound \eqref{EqnBound} is useful to put constraints on the conversion probability only if $\int_1^xdx'\,|D(x')|$ is bounded from above for all $x\geq1$.  As a side note, it is interesting to point out that the case $A(x)=B(x)$ in \eqref{EqnSolnAB} saturates the upper bound \eqref{EqnBound} if $D(x)\geq0\,\,\forall\,\,x$.

Finally, the form of the evolution equations \eqref{EqnEvol} leads to some qualitative understanding for magnetars in the simple case of a dipolar field $B=B_0(r_0/r)^3$ for which
\be 
A(x)-B(x)=[\Delta_a-\Delta_\parallel(x)]r_0=\left[\Delta_a-\frac{\Delta_{\parallel0}\hat{q}(x)}{x^6}\right]r_0,\qquad\qquad D(x)=\Delta_Mr_0=\frac{\Delta_{M0}r_0}{x^3}.\nonumber
\ee
First, since for a magnetar $D(x)\to0$ as $x\to\infty$, $\chi(x)$ is a constant independent of $x$ for $x$ large as expected from \eqref{EqnSolnC0}, \textit{i.e.} the conversion probability is a well-defined quantity with a fixed value far away from the magnetar.  This behavior should occur on physical ground since the magnetic field decreases as the state propagates away from the magnetar, leading to a vanishing mixing between ALPs and photons.

Moreover, from the upper bound \eqref{EqnBound} with a pure ALP initial state $\chi(1)=0$, the conversion probability is constrained by
\be
P_{a\to\gamma}(x)=\sin^2[\chi(x)]\leq\begin{cases}\sin^2\left[\int_1^xdx'\,D(x')\right]&\text{if }\int_1^xdx'\,D(x')<\frac{\pi}{2}\\1&\text{otherwise}\end{cases}.\nonumber
\ee
Hence, for a magnetar the conversion probability for a pure ALP initial state is constrained if $\int_1^\infty dx\,D(x)<\frac{\pi}{2}$.  In the simple case of a dipolar field, the conversion probability is smaller than $P_*=\sin^2(\Delta_{M0}r_0/2)$, \textit{i.e.} $P_{a\to\gamma}(\infty)\leq P_*$, if $\Delta_{M0}r_0<\pi$.  Unfortunately this bound is not relevant here since $\Delta_{M0}r_0\gg1$.  Indeed, for sample values such as $\omega=100\,\text{keV}$, $m_a=10^{-8}\,\text{keV}$, $g/e=10^{-15}\,\text{keV}^{-1}$, $r_0=10\,\text{km}$, $B_0=20\times10^{14}\,\text{G}$ and $\theta=\pi/2$, one obtains three rather different pure numbers controlling the solutions to the evolution equations given by $\Delta_ar_0\approx-2.5\times10^{-5}$, $\Delta_{\parallel0}r_0\approx8.6\times10^{13}$ and $\Delta_{M0}r_0\approx3.0\times10^5$ respectively, demonstrating that the analytic bound cannot be used in the following to constrain the ALP parameter space.\footnote{Here the plasma contribution is negligible for stellar magnetic fields and hard X-ray frequencies \cite{Raffelt:1987im}.}


\section{Results}\label{benchmark}

In this section, the evolution equations \eqref{EqnEvol} are solved numerically.  First, the dependence of the conversion probability on the ALP and magnetar parameters is shown for a benchmark point in the hard X-ray range.  Then, the conversion probability in the $(m_a,g)$ plane is given for several values of $\omega$.  The conversion probability and photon luminosity depend on properties of the magnetar, like its radius and magnetic field.  We choose the magnetic field $B_0=20\times10^{14}\,\text{G}$ corresponding to SGR 1806-20 and a typical magnetar radius $r_0=10\,\text{km}$ as our benchmark values.\footnote{Although the magnetic field is quite large, its position dependence is strong enough to ensure the weak-dispersion limit can be used where the conversion occurs, which is far away from the magnetar.}

\subsection{Dependence of the Conversion Probability}

For a magnetar in the simple case of a dipolar magnetic field, the conversion probability of a pure ALP initial state to a photon state, given by $P_{a\to\gamma}\equiv P_{a\to\gamma}(\infty)=\sin^2[\chi(\infty)]$, depends on six different parameters.  The ALP parameters are the ALP energy $\omega$, the ALP mass $m_a$ and the ALP-photon coupling constant $g$.  The magnetar parameters are the magnetar radius $r_0$, the (dimensionless) magnetar magnetic field at the surface $b_0=B_0/B_c$ and the angle between the direction of propagation and the magnetic field $\theta$.

\begin{figure}[!t]
\centering
\resizebox{15cm}{!}{
\includegraphics{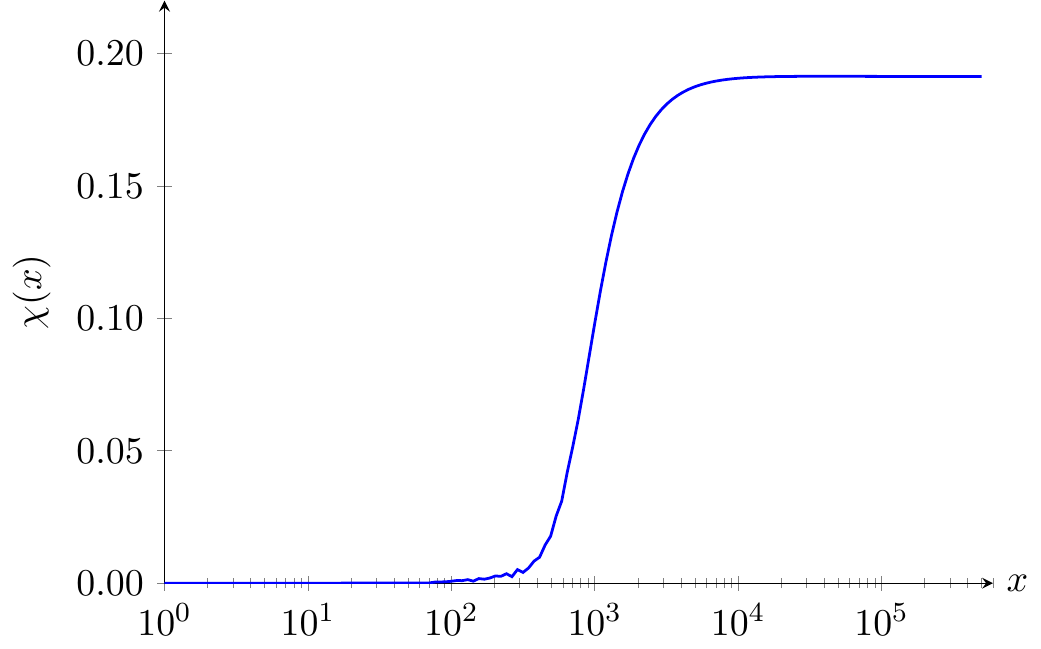}
\hspace{2cm}
\includegraphics{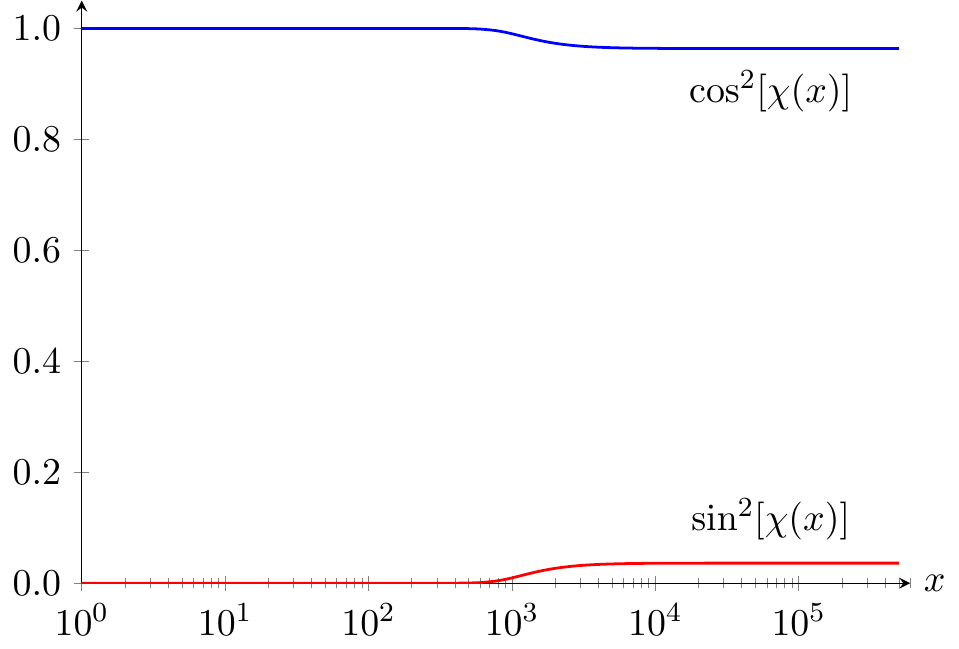}
}
\caption{Evolution of $\chi(x)$ (left panel) and $\cos^2[\chi(x)]$ and $\sin^2[\chi(x)]$ (right panel, blue and red curve respectively) as a function of the dimensionless distance $x$ from the magnetar surface for the benchmark point $\omega=100\,\text{keV}$, $m_a=10^{-8}\,\text{keV}$, $g/e=10^{-15}\,\text{keV}^{-1}$, $r_0=10\,\text{km}$, $B_0=20\times10^{14}\,\text{G}$ and $\theta=\pi/2$.}
\label{FigEvol}
\end{figure}
The evolution of $\chi(x)$ as well as $\cos^2[\chi(x)]$ and $\sin^2[\chi(x)]$ at the benchmark point $\omega=100\,\text{keV}$, $m_a=10^{-8}\,\text{keV}$, $g/e=10^{-15}\,\text{keV}^{-1}$, $r_0=10\,\text{km}$, $B_0=20\times10^{14}\,\text{G}$ and $\theta=\pi/2$ is shown in Fig.~\ref{FigEvol}.  The benchmark point is chosen in the hard X-ray range with appropriate magnetar parameters for SGR 1806-20.

\begin{figure}[!t]
\centering
\resizebox{15cm}{!}{
\includegraphics{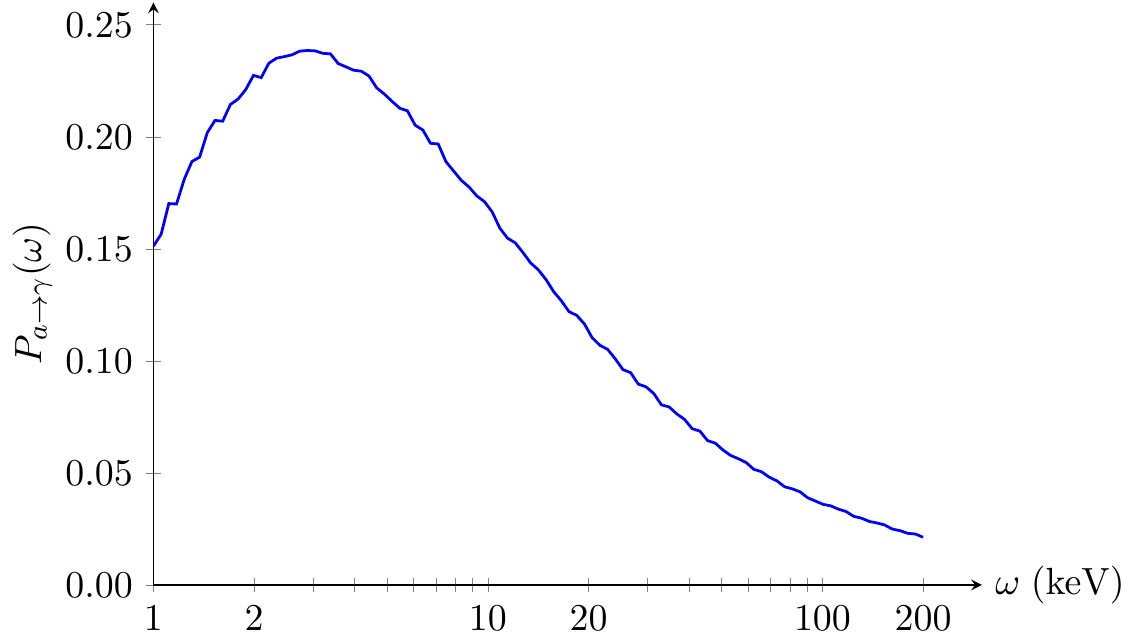}
\hspace{2cm}
\includegraphics{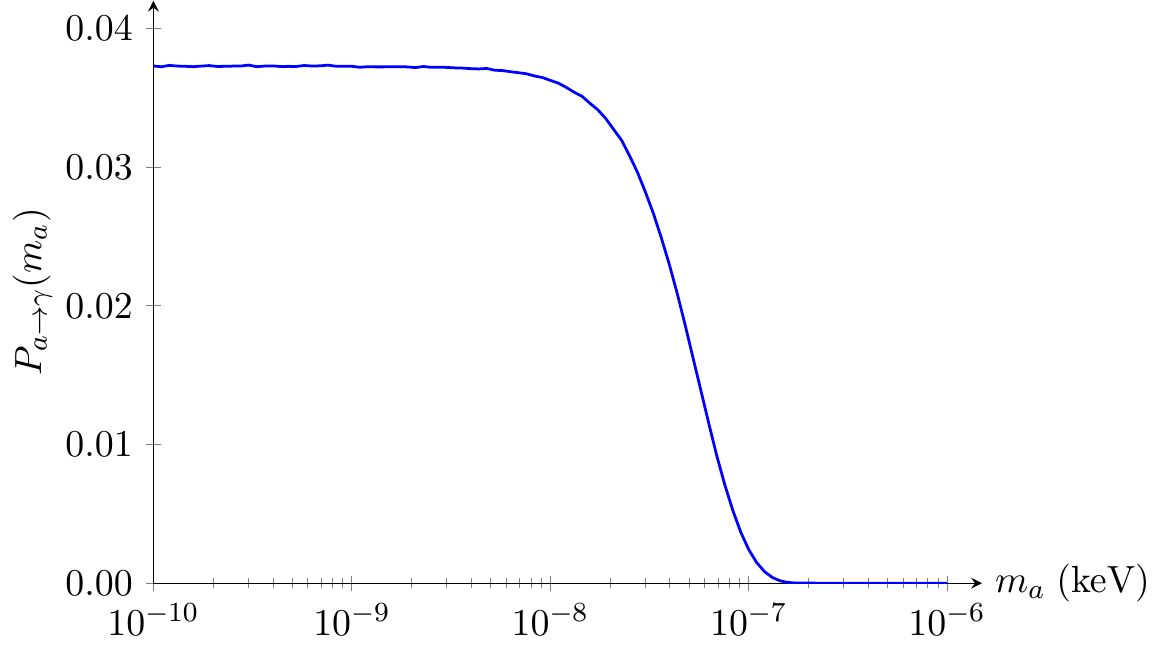}
}
\resizebox{15cm}{!}{
\includegraphics{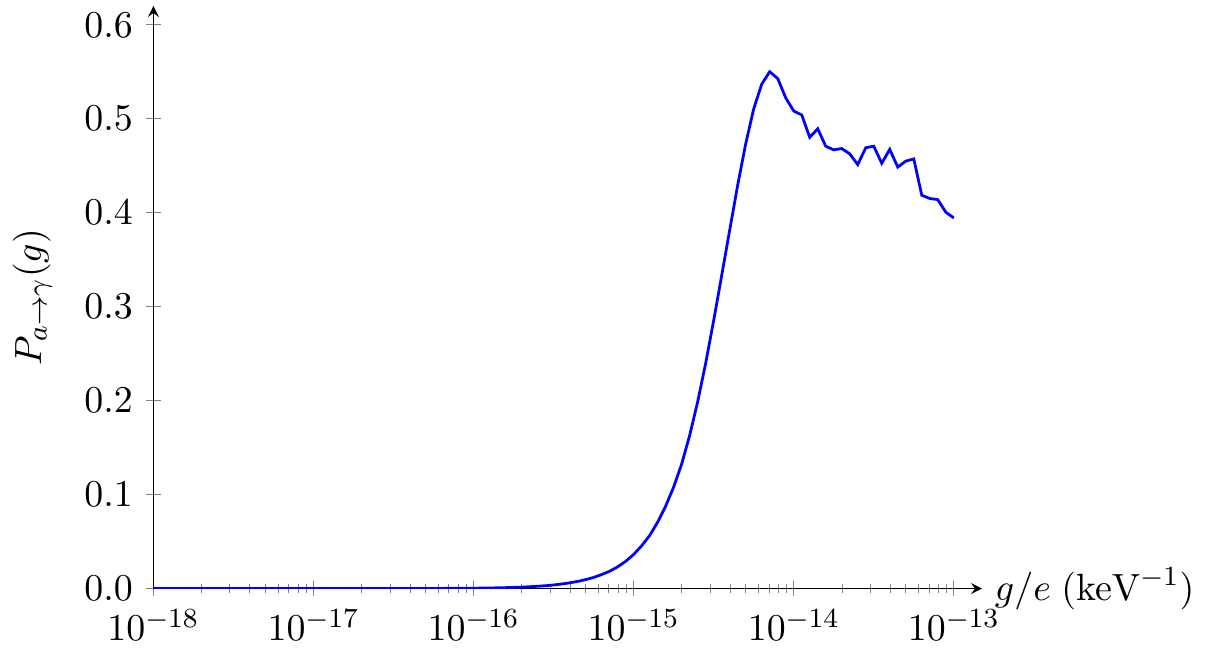}
\hspace{2cm}
\includegraphics{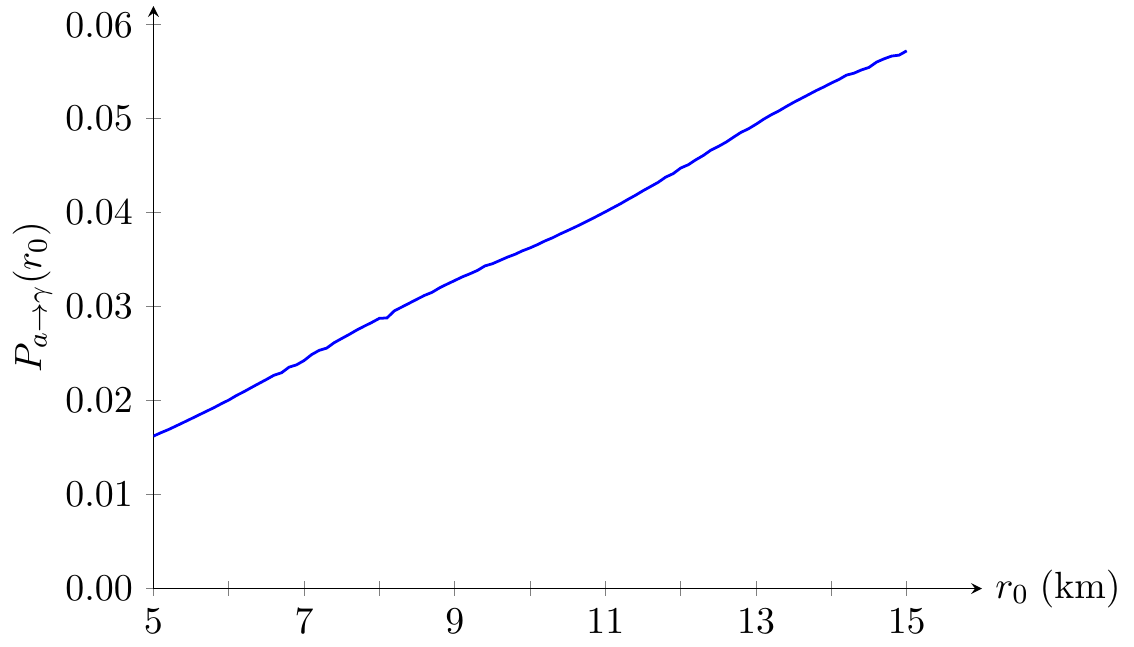}
}
\resizebox{15cm}{!}{
\includegraphics{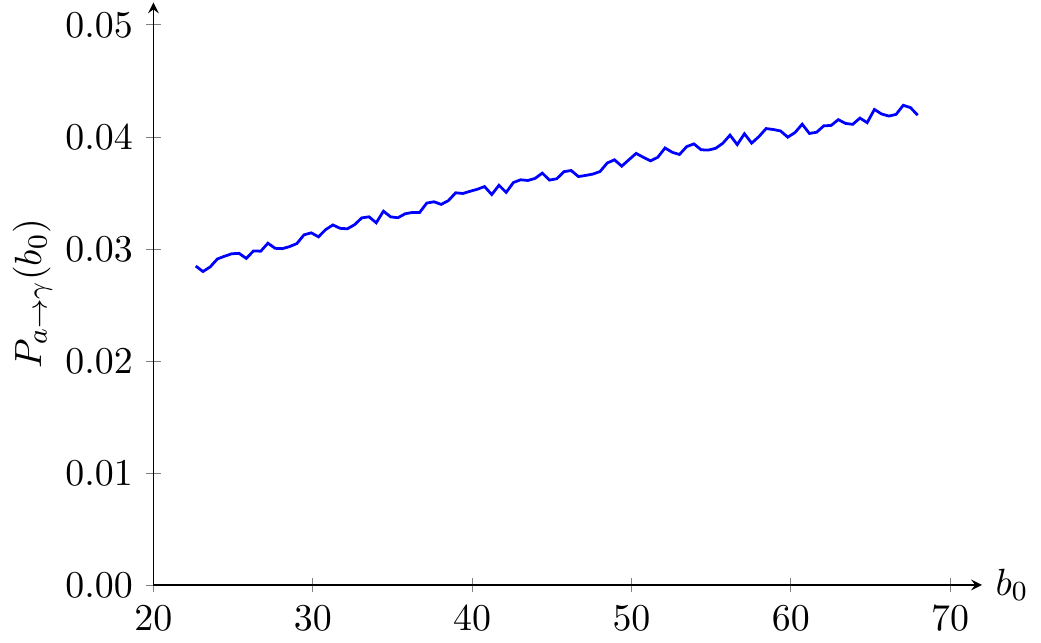}
\hspace{2cm}
\includegraphics{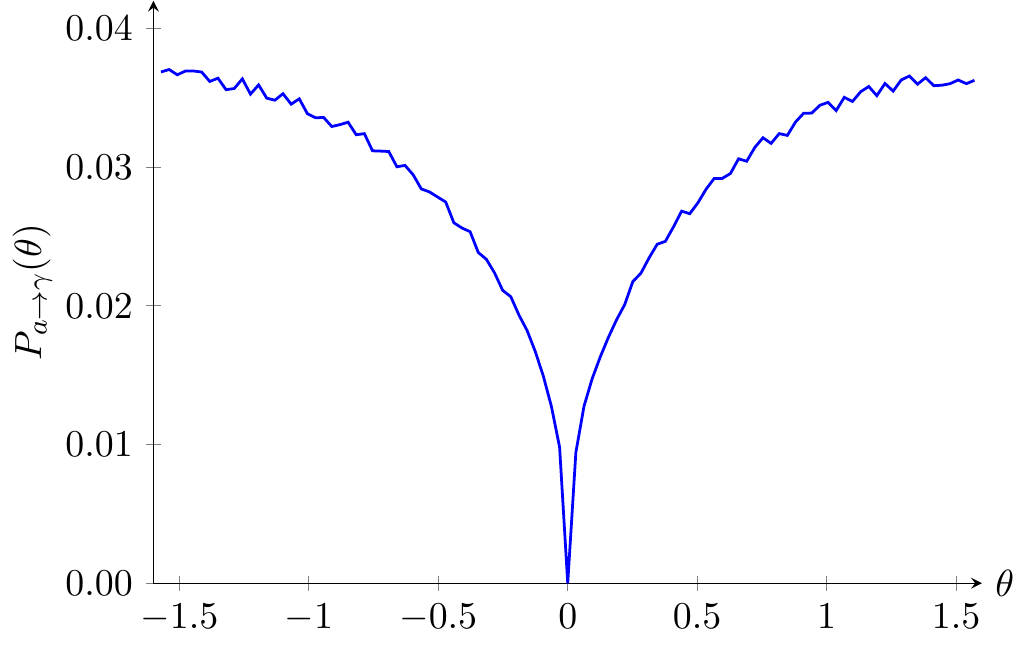}
}
\caption{ALP-to-photon conversion probability in function of the ALP and magnetar parameters for the benchmark point $\omega=100\,\text{keV}$, $m_a=10^{-8}\,\text{keV}$, $g/e=10^{-15}\,\text{keV}^{-1}$, $r_0=10\,\text{km}$, $B_0=20\times10^{14}\,\text{G}$ and $\theta=\pi/2$.  Each panel shows the dependence of the conversion probability around the benchmark point for one of the ALP or magnetar parameter.}
\label{FigBP}
\end{figure}
The conversion probability $P_{a\to\gamma}=\sin^2[\chi(\infty)]$ around the same benchmark point in function of one of the ALP or magnetar parameter is shown in the corresponding panel of Fig.~\ref{FigBP}.

Several comments are in order for this benchmark point.  First, the conversion probability peaks in function of the ALP energy $\omega$ in the X-ray range.  This observation is important since the normalized ALP spectrum from nucleon-nucleon bremsstrahlung emission for a degenerate medium relevant to magnetars peaks in the hard X-ray range for our benchmark model.  Second, with respect to the ALP mass $m_a$, the conversion probability plateaus around a non-zero (zero) value for small (large) ALP masses, with a sharp transition between the two regimes at around $m_a\approx5\times10^{-8}\,\text{keV}$.  Hence the conversion probability vanishes for large ALP masses.  This behavior can be explained qualitatively from \eqref{EqnEvol}.  Indeed, since $D(x)\sim1/x^3$, the conversion probability reaches a well-defined limit as $x$ increases.  Beyond some distance $x_*$ from the magnetar, the conversion probability is essentially fixed.  On the one hand, if the ALP mass is small enough such that $|A(x_*)|\ll|D(x_*)|$, then $|A(x)|\ll|D(x)|$ $\forall$ $x\in(1,x_*)$ and thus the evolution equations from the surface to $x_*$ only have a negligible dependence on the ALP mass.  On the other hand, if the ALP mass is large enough such that $|A(1)|\gg|D(1)|$, then $|A(x)|\gg|D(x)|$ $\forall$ $x\in(1,\infty)$.  Since $A(x)-B(x)$ has a definite sign, $|A(x)-B(x)|\gg|D(x)|$ and $\Delta\phi(x)$ decreases rapidly in the evolution equations, leading to variations on $\chi(x)$ that average out, implying a vanishing conversion probability for large ALP masses.  Third, the dependence on the ALP-photon coupling constant $g$ is important, as the conversion probability is zero for $g/e\lesssim2\times10^{-16}\,\text{keV}^{-1}$ and increases significantly for larger $g$.  The vanishing conversion probability for small $g$ is expected since ALP-photon oscillations are suppressed as $g$ decreases.  Effects of the oscillatory nature of the problem can also be seen for larger $g$.  Fourth and fifth, a change of the magnetar radius $r_0$ or the dimensionless surface magnetic field $b_0$ leads to a variation of the conversion probability that is quite mild.  Sixth, the dependence of the conversion probability with respect to the angle $\theta$ is also mild away from $\theta=0$ where it vanishes, as expected from the ALP-photon coupling $\Delta_M$.  Moreover, the conversion probability is an even function with respect to the angle $\theta$, \textit{i.e.} $P_{a\to\gamma}(-\theta)=P_{a\to\gamma}(\theta)$.  This is expected since a solution to the evolution equations \eqref{EqnEvol} with $\chi(1)=0$, $\Delta\phi(1)=0$ and $\theta$ is also a solution to the evolution equations with $\chi(1)=0$, $\Delta\phi(1)=\pi$ and $-\theta$.  Since there is no physical difference between $\Delta\phi(1)=0$ and $\Delta\phi(1)=\pi$, the conversion probability far away from the magnetar should not depend heavily on this particular choice.\footnote{The choice $\Delta\phi(1)=\pi$ is as physically motivated as our initial choice of $\Delta\phi(1)=0$.}

\begin{figure}[!t]
\centering
\resizebox{15cm}{!}{
\includegraphics{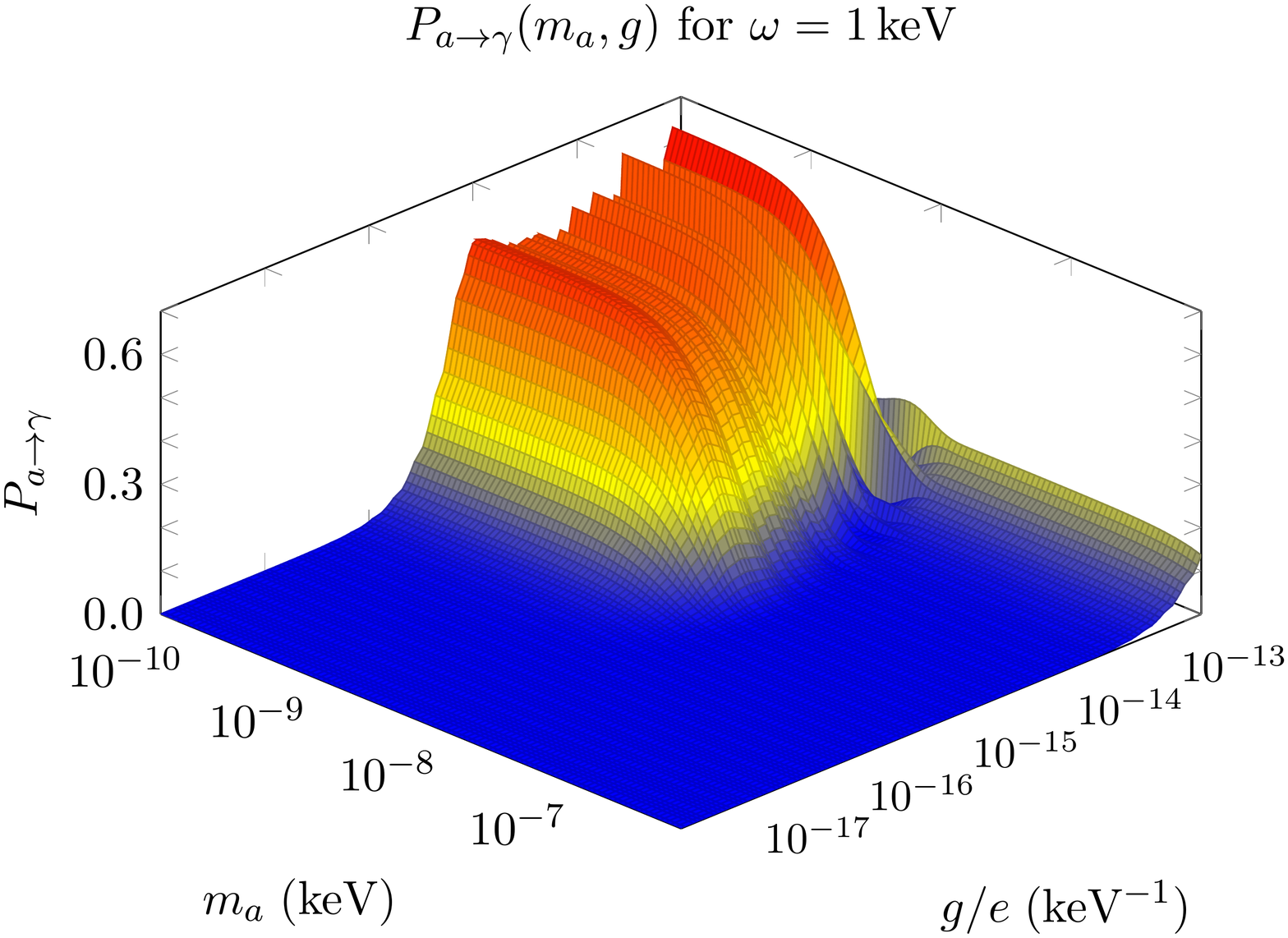}
\hspace{2cm}
\includegraphics{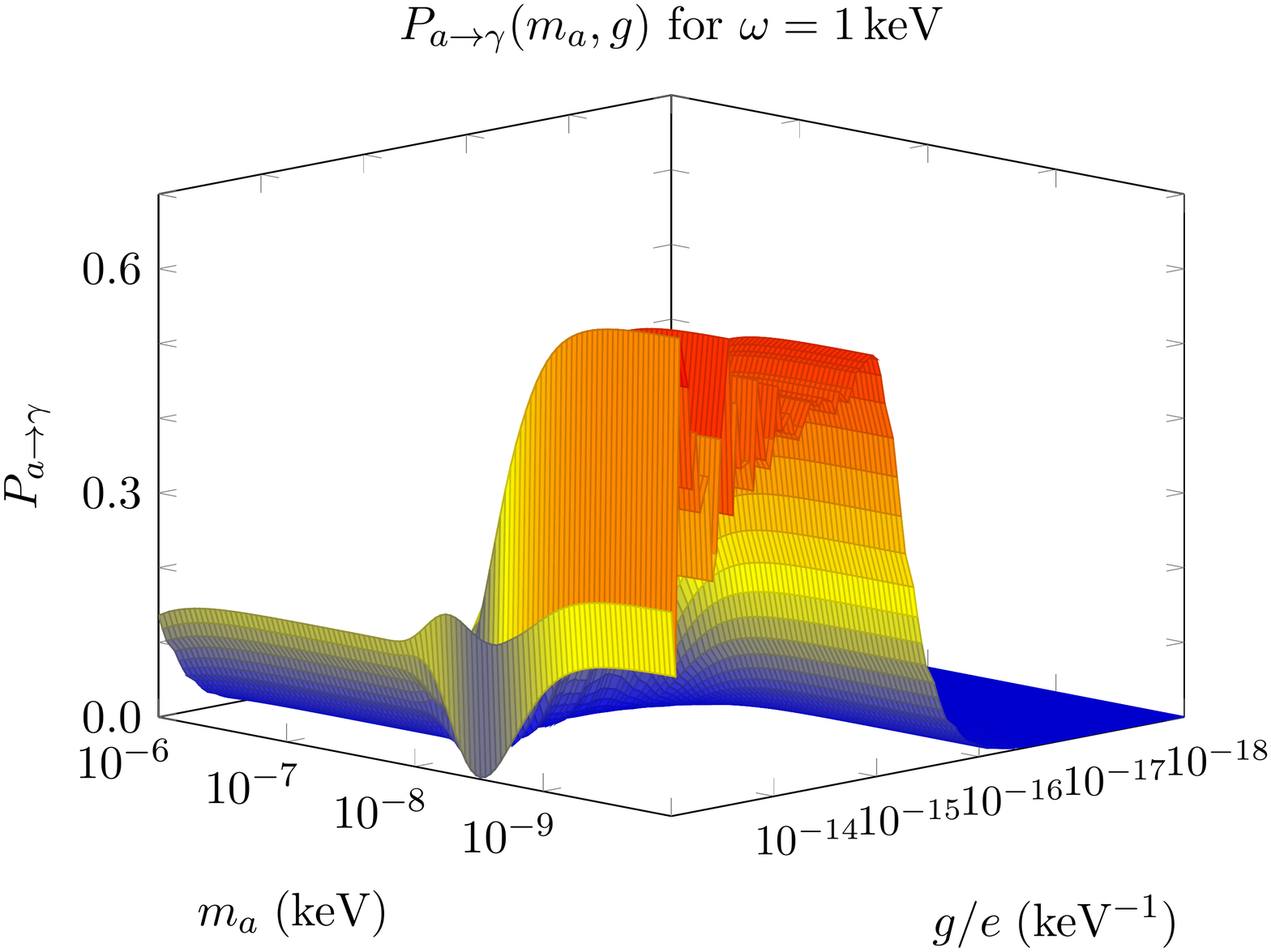}
}
\resizebox{15cm}{!}{
\includegraphics{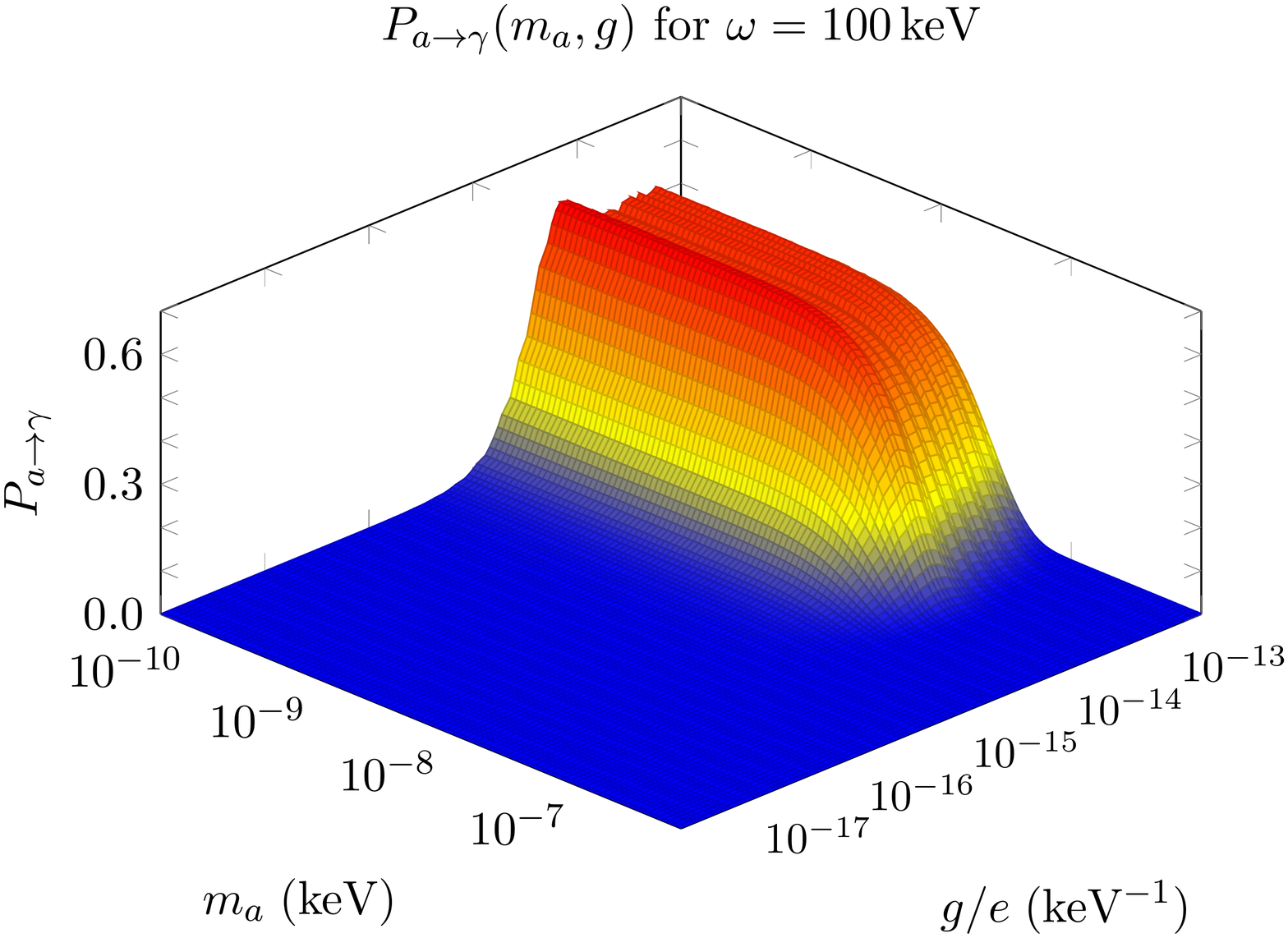}
\hspace{2cm}
\includegraphics{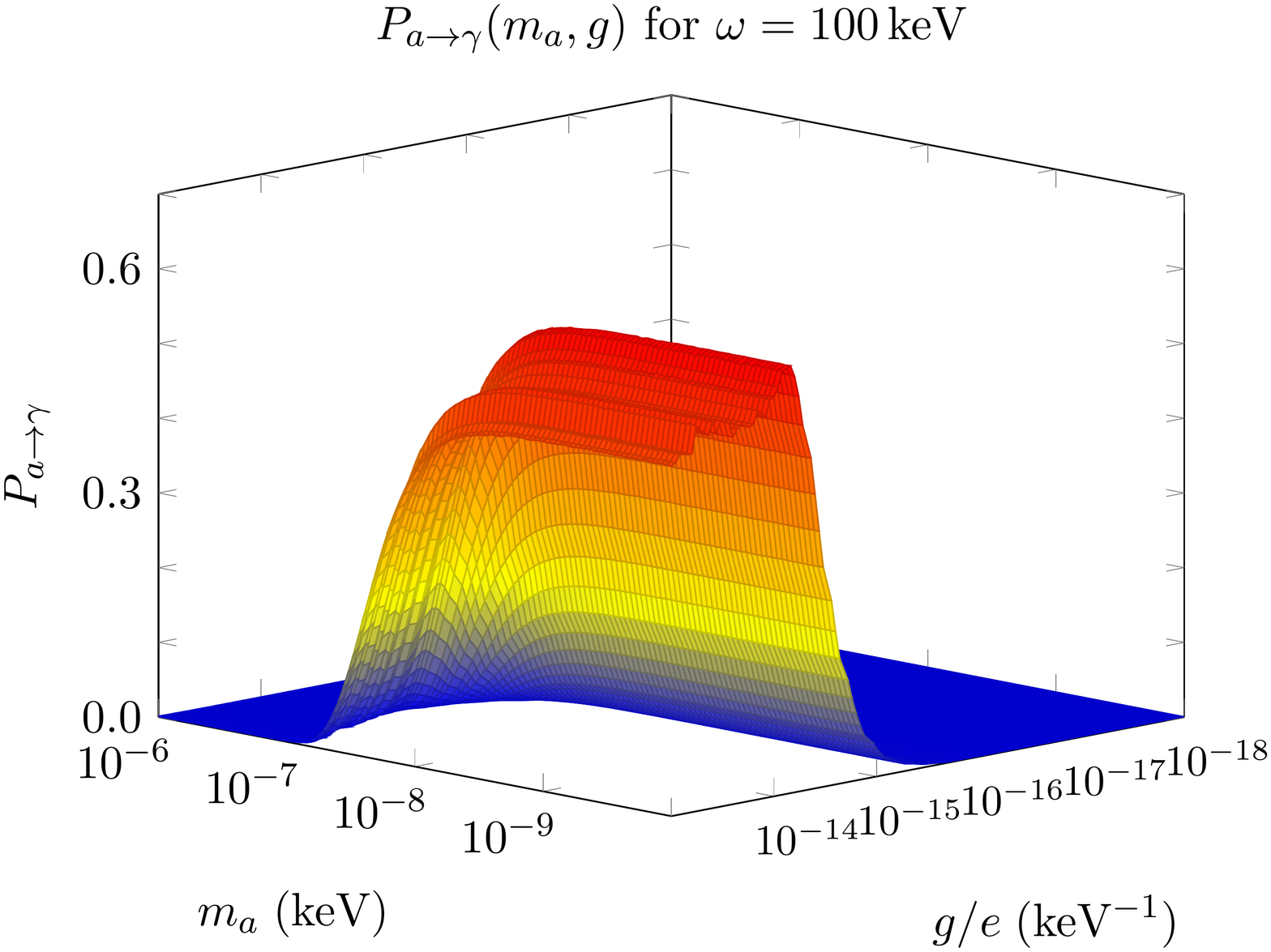}
}
\resizebox{15cm}{!}{
\includegraphics{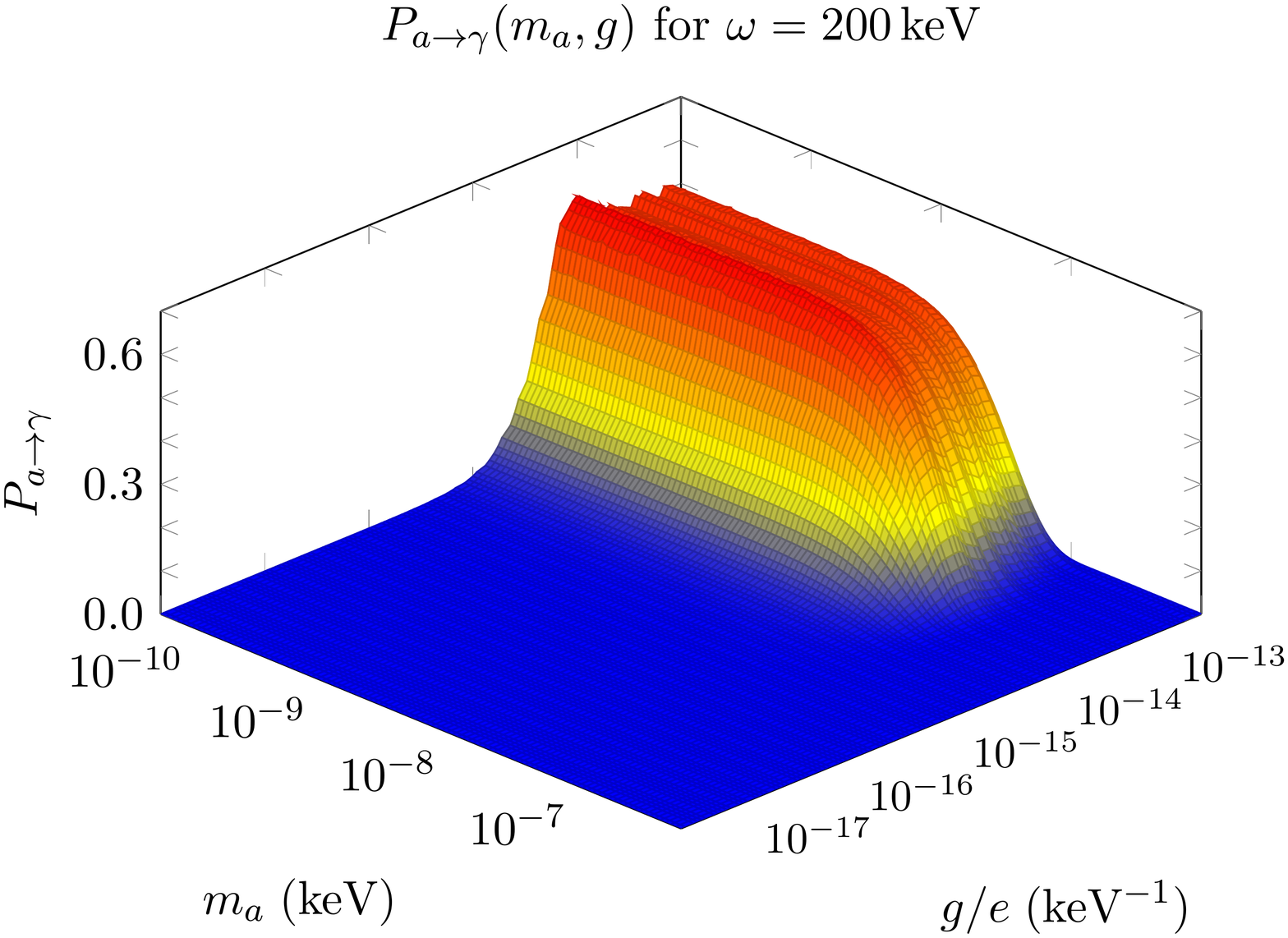}
\hspace{2cm}
\includegraphics{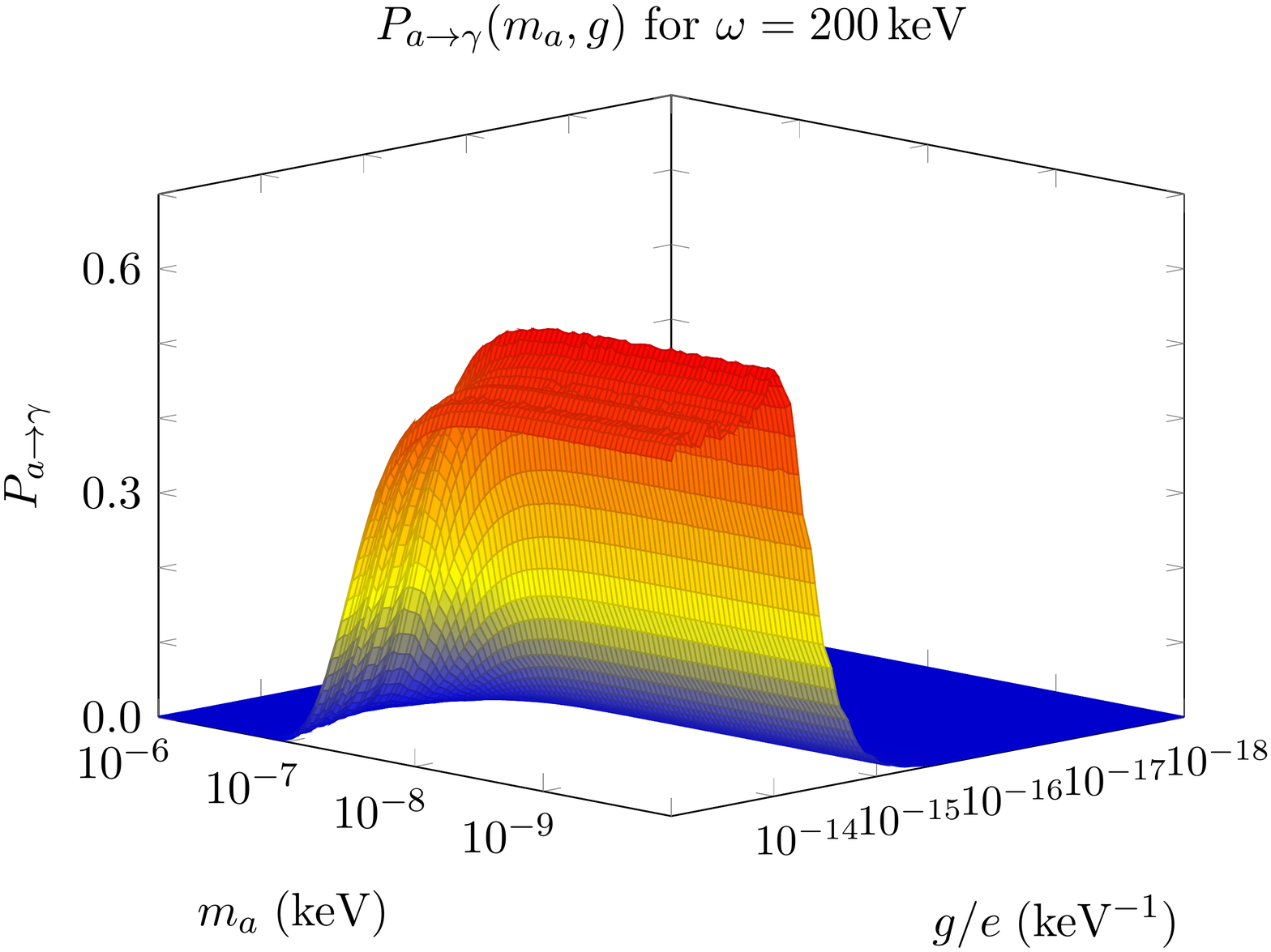}
}
\caption{ALP-to-photon conversion probability in the $(m_a,g)$ plane for different values of the ALP energy $\omega$ corresponding to $\omega=1\,\text{keV}$ (top panels), $\omega=100\,\text{keV}$ (middle panels) and $\omega=200\,\text{keV}$ (bottom panels).  The two panels (left and right) show the same conversion probability from different points of view.}
\label{FigCP}
\end{figure}
The conversion probability in the $(m_a,g)$ plane for $\omega=1$, $100$ and $200\,\text{keV}$ with magnetar parameters $r_0=10\,\text{km}$ and $B_0=20\times10^{14}\,\text{G}$ relevant to our benchmark model and $\theta=\pi/2$ is shown in Fig.~\ref{FigCP}.

Again, several comments are in order.  Fig.~\ref{FigCP} shows that as the ALP energy $\omega$ increases, the locations of the transitions in $m_a$ and $g$ both increase.  Moreover, the conversion probability is negligible everywhere except for small ALP mass and large ALP-photon coupling constant, as expected from our previous discussion.  In the latter region, the oscillatory nature of the problem can clearly be seen.\footnote{Moreover, contrary to \cite{Raffelt:1987im}, $P_{a\to\gamma}$ can be large since $\Delta_ar_0$, $\Delta_\parallel r_0$ and $\Delta_Mr_0$ depend differently on $x$ \cite{Lai:2006af}.}

With the $\omega$- and $\theta$-dependence of the conversion probability in the $(m_a,g)$ plane, it is now straightforward to compute the ALP-to-photon luminosity for SGR 1806-20, assuming $r_0=10\,\text{km}$ and $B_0=20\times10^{14}\,\text{G}$.  From Fig.~\ref{FigCP}, we therefore expect that the constraint on the magnetar photon luminosity in the hard X-ray range will exclude the ALP parameter space with small ALP mass and large ALP-photon coupling constant.

\subsection{Photon Luminosity}

The photon luminosity from ALP-photon oscillations in the band $\omega\subset(\omega_i,\omega_f)$ can be computed from the ALP-to-photon conversion probability obtained above.

Indeed, from the normalized ALP spectrum $dN_a/d\omega$ such that $\int_0^\infty d\omega\,dN_a/d\omega=1$, the ALP-to-photon luminosity in the band $\omega\subset(\omega_i,\omega_f)$ is given by
\be\label{EqnLum}
L_{a\to\gamma}=\frac{N_a}{2\pi}\int_0^{2\pi}d\theta\int_{\omega_i}^{\omega_f}d\omega\,\omega\frac{dN_a}{d\omega}P_{a\to\gamma}(\omega,\theta),
\ee
where $N_a$ is the total number of ALPs emitted by the magnetar.  The $\theta$-average in \eqref{EqnLum} is necessary to obtain the magnetar luminosity.  However, since the $\theta$-average is computationnally intensive, it is replaced by
\be
\frac{1}{2\pi}\int_0^{2\pi}d\theta\,P_{a\to\gamma}(\omega,\theta)\to R_\theta P_{a\to\gamma}(\omega,\pi/2),\nonumber
\ee
where $R_\theta$ is a conservative suppression factor computed for several $(\omega,m_a,g)$ points which is numerically given by $R_\theta=0.6$.

The total number of ALPs emitted by the magnetar is constrained by cooling models.  Indeed, since magnetar cooling is well understood in terms of neutrino cooling luminosity, the ALP cooling luminosity should not overtake the neutrino cooling luminosity, hence
\be\label{EqnNa}
L_a=N_a\int_0^\infty d\omega\,\omega\frac{dN_a}{d\omega}\leq L_\nu=4\pi\int_0^{r_0} dr\,r^2\dot{q}_\nu,
\ee
where $\dot{q}_\nu$ is the neutrino emissivity.  Assuming that the neutrino emissivity is constant throughout the whole magnetar, \eqref{EqnNa} implies
\be
N_a\leq\frac{4\pi r_0^3\dot{q}_\nu}{3\int_0^\infty d\omega\,\omega\frac{dN_a}{d\omega}},\nonumber
\ee
and the ALP-to-photon luminosity \eqref{EqnLum} which saturates the previous bound is\footnote{As explained in the appendix, for specific models with explicit dependence on $g_{aN}$, the luminosity $L_{a\to\gamma}$ in \eqref{EqnLumB} must be replaced by $(g_{aN}/g_{aN}^*)^2L_{a\to\gamma}$ where $g_{aN}^*$ is given by \eqref{EqngaN}.}
\be\label{EqnLumB}
L_{a\to\gamma}=\frac{4\pi r_0^3\dot{q}_\nu R_\theta}{3\int_0^\infty d\omega\,\omega\frac{dN_a}{d\omega}}\int_{\omega_i}^{\omega_f}d\omega\,\omega\frac{dN_a}{d\omega}P_{a\to\gamma}(\omega,\pi/2).
\ee
In the following, the constraints on the ALP parameter space are computed from the ALP-to-photon luminosity \eqref{EqnLumB}.

Usually, it is necessary to make some assumptions on the ALP production mechanisms as well as the ALP-nucleon coupling constant $g_{aN}$ to determine the normalized ALP spectrum and the total number of ALPs emitted by the magnetar.  However, to obtain conservative constraints on the ALP parameter space, we choose to take the normalized ALP spectrum \eqref{EqnBrem} from nucleon-nucleon bremsstrahlung emission for a degenerate medium relevant to magnetars \cite{Raffelt:1987im}.  Moreover, as we have done above, we bound the total number of ALPs emitted by the magnetar with respect to the neutrino cooling luminosity, but using modified URCA emission \eqref{EqnURCA} \cite{Friman:1978zq}.  The nucleon-nucleon bremsstrahlung emission of ALPs and modified URCA emission of neutrinos are the respective dominant production mechanisms and their use is justified to stay as conservative as possible and to impose as few assumptions as possible on the ALP and neutrino production mechanisms.  Both production mechanisms are discussed at greater length in the appendix.


\section{Constraints on ALP Parameter Space}\label{SSGR}

In this section, we put together all the ingredients to compute the total ALP-to-photon lumnosity.  We then compare our results with observations. 

\subsection{Photon Luminosity}

For the conversion probability, we choose the benchmark values of $r_0$ and $B_0$ used in the rest of the paper.  The ALP emissivity, on the other hand, is bounded by the neutrino cooling luminosity via modified URCA emission.  This is strongly dependent on the core temperature of the magnetar. 

\begin{figure}[!t]
\centering
\resizebox{15cm}{!}{
\includegraphics{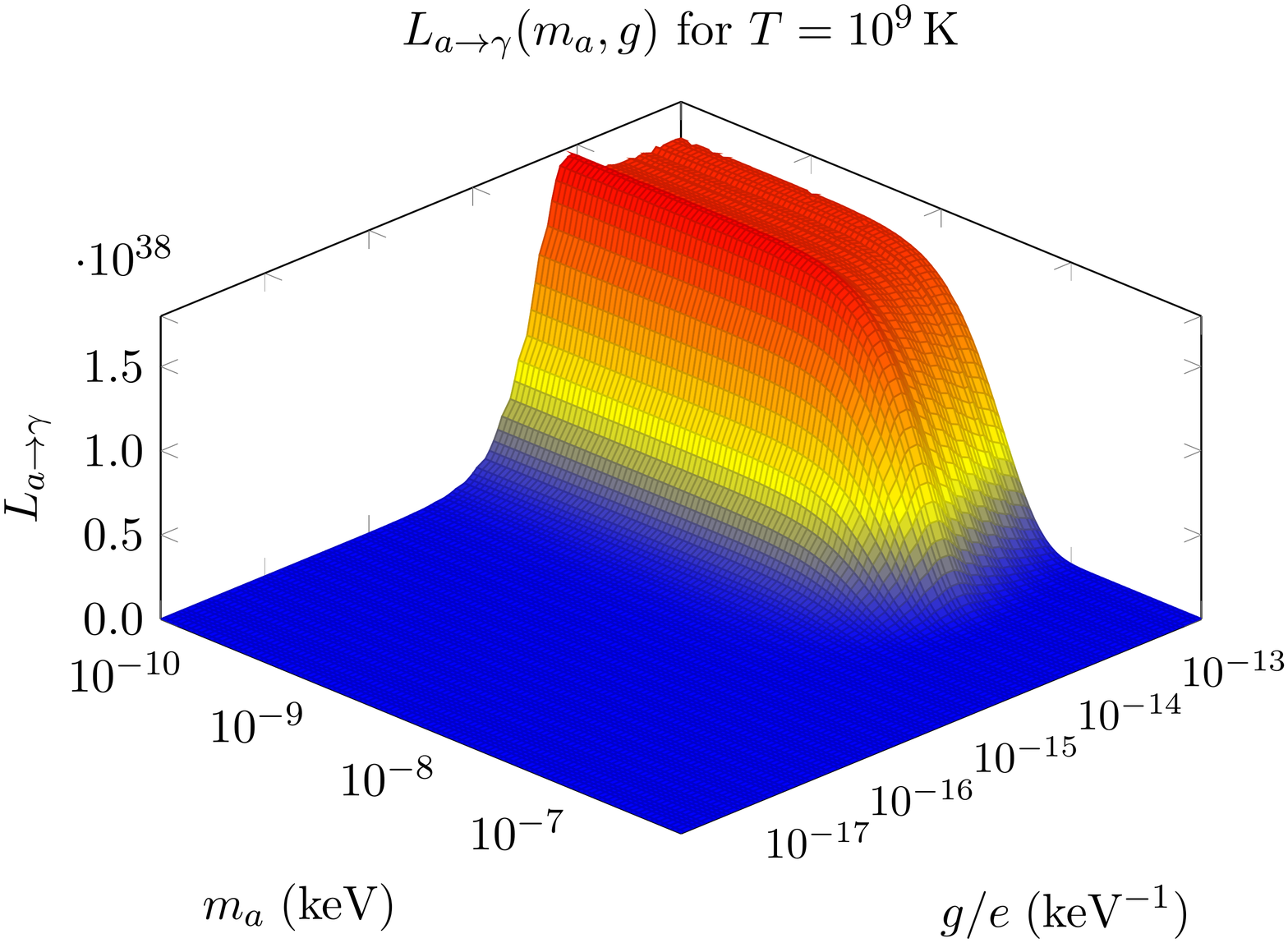}
\hspace{2cm}
\includegraphics{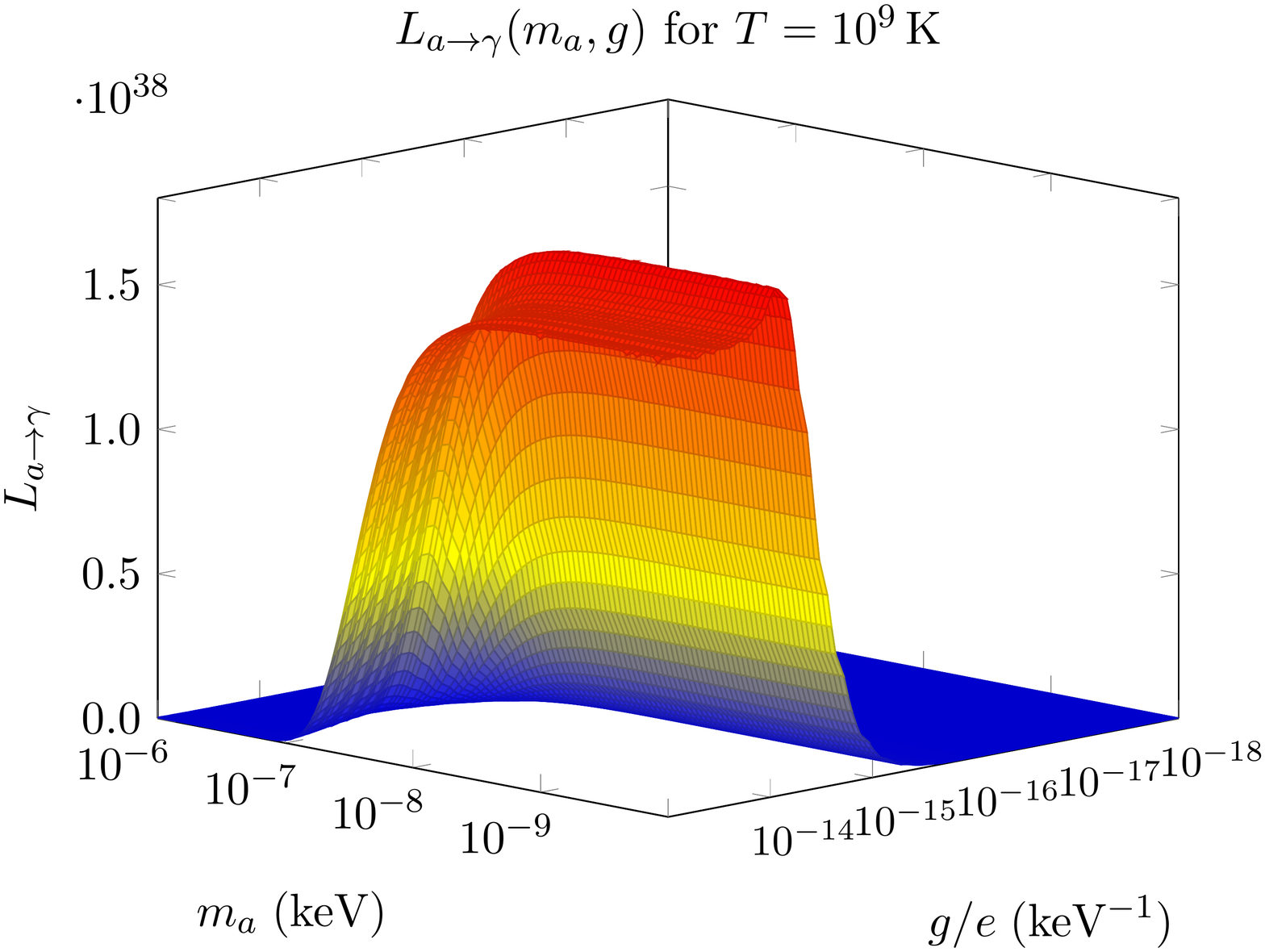}
}
\caption{Photon luminosity from ALP-photon oscillations in the broad band from $1\,\text{keV}$ to $200\,\text{keV}$ in the $(m_a,g)$ plane.  The computations are done for SGR 1806-20 assuming $r_0=10\,\text{km}$ and $B_0=20\times10^{14}\,\text{G}$.  The magnetar core temperature is assumed to be $T=10^9\,\text{K}$.  The two panels show the same conversion probability from different points of view.}
\label{FigPL}
\end{figure}
In Fig.~\ref{FigPL}, we first display the photon luminosity in the broad band from $\omega_i=1\,\text{keV}$ to $\omega_f=200\,\text{keV}$  assuming an ALP emissivity that equals the modified URCA emission of neutrinos for a core temperature of $T=10^9\,\text{K}$.  The photon luminosity is obtained from \eqref{EqnLumB}, \eqref{EqnURCA}, \eqref{EqnBrem} and the conversion probability.  The results are displayed on the $(m_a,g)$ plane. 

\subsection{ALP Parameter Space Constraints}

While the surface temperature of the magnetar can be easily deduced from the thermal emission, the relation between the surface and core temperatures depends on a variety of factors that affect the conduction of heat.  These include the strength of the magnetic field in the blanketing envelope and its angle with respect to the radial direction, as well as the chemical composition of the magnetar.  An exploration of these effects is beyond the scope of our paper.  We refer to \cite{Beloborodov:2016mmx} and references therein for a thorough discussion.

\begin{figure}[!t]
\centering
\resizebox{9cm}{!}{
\includegraphics{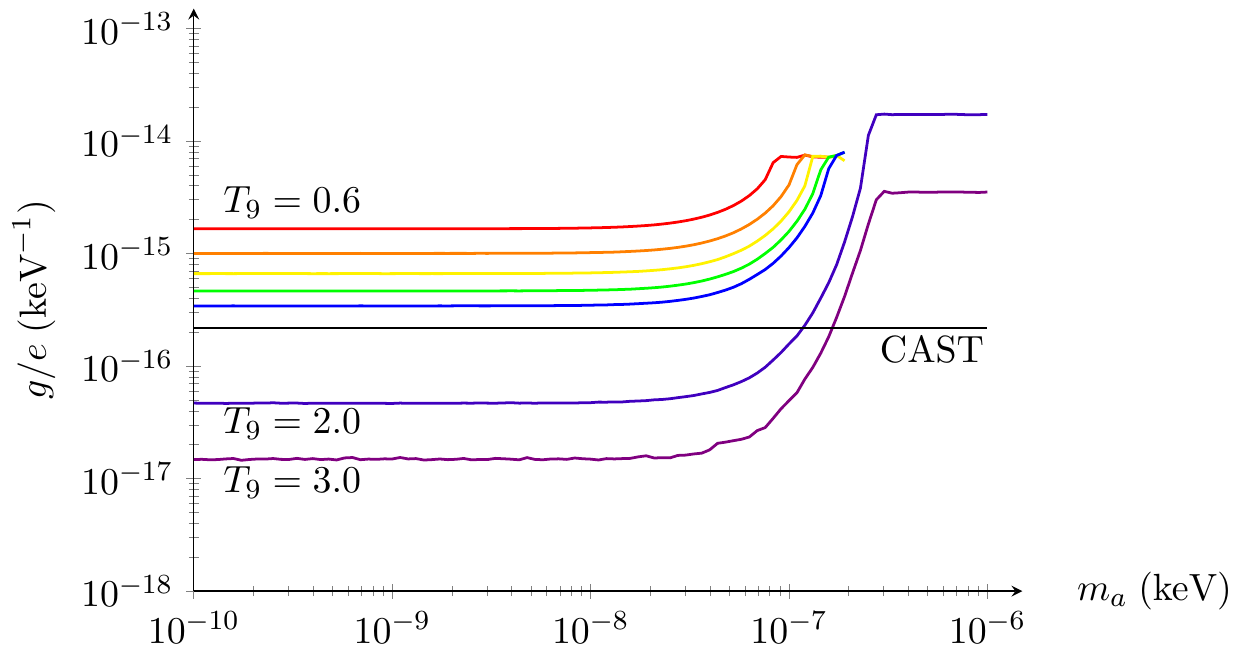}
}
\caption{Exclusion contours in the $(m_a,g)$ plane for different magnetar core temperatures.  For a given magnetar core temperature, all the ALP parameter space above the corresponding curve is excluded.  The computations are done for SGR 1806-20 assuming $r_0=10\,\text{km}$ and $B_0=20\times10^{14}\,\text{G}$.  The magnetar core temperatures span the range from $T_9=0.6$ to $T_9=3.0$ where $T_9=T/(10^9\,\text{K})$.  From top to bottom, the curves correspond to $T_9=0.6$, $0.7$, $0.8$, $0.9$, $1.0$, $2.0$ and $3.0$ respectively (\textit{i.e.} the red, orange, yellow, green, blue, indigo and violet curve).  For comparison, the exclusion contour from CAST is shown in black.}
\label{FigExcl}
\end{figure}

We instead display our results for several core temperatures between $T=6\times10^8\,\text{K}$ and $T=3\times10^9\,\text{K}$.  Since the observed luminosity of SGR 1806-20 in this range is $L_\gamma^\text{obs}=1.2\times10^{36}\,\text{erg}\cdot\text{s}^{-1}$ \cite{Molkov:2004sy,Kaspi:2017fwg},\footnote{The observed luminosity in the hard X-ray band in \cite{Molkov:2004sy} is quoted as $L_\gamma^\text{obs}=3.6\times10^{36}\,\text{erg}\cdot\text{s}^{-1}$ for an assumed distance of $15\,\text{kpc}$.  Since the distance is now believed to be $8.7\,\text{kpc}$ \cite{Kaspi:2017fwg}, we modified the observed luminosity accordingly.} any point in the $(m_a,g)$ plane with $L_{a\to\gamma}>L_\gamma^\text{obs}$ is excluded.  The exclusion curves, where $L_{a\to\gamma}=L_\gamma^\text{obs}$, are shown in Fig.~\ref{FigExcl}.
As expected, the excluded region in the ALP parameter space corresponds to small ALP mass and large ALP-photon coupling constant.  Furthermore, the excluded region is larger for higher magnetar core temperature.  In fact, due to the strong neutrino emissivity dependence on the magnetar core temperature, the largest total number of ALPs emitted by the magnetar allowed by the cooling argument also has a strong dependence on the magnetar core temperature, which translates into exclusion contours in the $(m_a,g)$ plane even where the conversion probability is negligible for high core temperature.

For comparison, the exclusion contour from the CAST helioscope experiment \cite{Anastassopoulos:2017ftl}, based entirely on the Primakoff process, is also shown.  From Fig.~\ref{FigExcl}, it is clear that the magnetar constraints on the ALP parameter space are better than CAST only for high magnetar core temperatures $T\gtrsim1\times10^9\,\text{K}$.

As mentioned before, it is also possible to use an opposite point of view.  Indeed, since the mechanism responsible for hard X-ray quiescent emission in magnetars is not known, from the analysis presented here one can argue that ALPs exist and are produced in magnetars with a subdominant luminosity to neutrinos such that magnetar cooling is not disturbed.  With magnetar core temperatures that satisfy $T\gtrsim1\times10^9\,\text{K}$ in line with the magnetar model \cite{Duncan:1992hi}, the magnetar hard X-ray emission (possibly with the appropriate spectral feature) could be generated by ALP-to-photon conversion in the magnetosphere without violating the bound from CAST.  A detailed analysis of all important production mechanisms for hard X-ray photons and axions must however be undertaken and the evolution equations must then be solved with the appropriate initial conditions (for example, with a mixed initial state if the amplitudes are comparable) before such conclusions can be reached.\footnote{It is important to note that the ALP mass and ALP-photon coupling would not be determined from this point of view, they would only have to satisfy the appropriate inequalities.  A spectral analysis could zero in on the right ALP parameters.}


\section{Conclusion}

Our goal in this paper has been to exploit the rapidly advancing field of magnetar science to study the physics of ALPs.  Magnetars, with their extremely strong magnetic fields, form a natural arena for investigating ALPs.

Our basic idea was to consider the conversion of ALPs emitted from the core of the neutron star into photons in the magnetosphere.  We assumed that the emission rate for ALPs is just subdominant to the neutrino emission rate for a given temperature.  For nucleon-nucleon bremsstrahlung, we obtain a broad ALP spectrum peaked around $\omega\sim3.3T$.  The coupled differential equations describing ALP-photon propagation in the magnetosphere were converted into a form that is efficient for extensive scans over multiple parameters.  We then presented the conversion probability as a function of the ALP energy, mass, coupling $g$, surface magnetic field strength $B_0$ of the magnetar, magnetar radius $r_0$, and angle between the magnetic field and the direction of propagation $\theta$.  Along the way, we developed an analytic formalism to perform similar calculations in more general $n$-state oscillation systems.

Taking benchmark values of the radius, magnetic field, and core temperature of SGR 1806-20, we then constrained the ALP-photon coupling by requiring that the photon flux coming from ALP conversion cannot exceed the observed luminosity of the magnetar.  Our results are depicted in Fig.~\ref{FigExcl}.

There are several future directions that would be interesting to explore.  Firstly, our approach has been to consider the photon flux from ALP-to-photon conversion for the entire energy range between $\omega=1-200\,\text{keV}$, and compare that to the broad band spectrum of the quiescent emission from SGR 1806-20 in the same range.  It would be interesting to perform a bin-by-bin spectral analysis, and presumably the constraints one would obtain from such an analysis would be more stringent.  Secondly, a polarization analysis along the lines of \cite{Lai:2006af,Perna:2012wn}, but in the hard X-ray band, would be very interesting.  Another aspect of our work that merits further study is the incorporation of other ALP production mechanisms -- such as electron bremsstrahlung on the surface -- and their relation to ALP-photon conversion.  Finally, our analytical treatment of the ALP-photon conversion probability can be utilized in other contexts, apart from magnetar physics, for example, in extra-galactic ALP-photon conversion (\cite{Kartavtsev:2016doq} and references therein).

We make a few comments about the observational and astrophysical aspects that affect our analysis.  The core temperature is clearly the parameter that most strongly influences our results, and we refer to Section 2 of \cite{Beloborodov:2016mmx} and references therein for a discussion of the relevant astrophysical modelling.  Our results also depend on the radius [through \eqref{EqnLumB}] and mass [through \eqref{EqnURCA}] of the magnetar, for which we took standard benchmark values.  We refer to the recent review \cite{Ozel:2016oaf} for observational prospects of the mass-radius relation and equation of state.  The observed luminosity in the hard X-ray regime also significantly affects the limits presented in Fig.~\ref{FigExcl}.  Results from the Hard X-ray Modulation Telescope (HXMT) will be very useful in further understanding the mechanism outlined in our paper.

We conclude with an intriguing speculation.  In recent years, satellites like INTEGRAL, RXTE, XMM-Newton, ASCA and NuSTAR have revealed that a considerable fraction of the bolometric luminosity of magnetars falls in the hard, rather than the soft, X-ray band \cite{Cavallari:2017rcq}.  While this has been observed for around nine magnetars, it is difficult to rule out this phenomenon for non-detected sources \cite{Turolla:2015mwa}.  The process of hard X-ray emission considered in this paper -- ALP production from the core followed by conversion in the magnetosphere -- produces a spectral peak in the correct range, and could be making an appreciable contribution to the observed luminosity.


\begin{center}
\textbf{Acknowledgements}
\end{center}

\noindent JFF is supported by NSERC and FRQNT.  KS would like to thank Matthew Baring for a very illuminating discussion on the current work, and possible future directions.  He would also like to thank Eddie Baron for useful discussions.


\appendix

\section{Neutrino and ALP Production Mechanisms}

This appendix discusses the simplest production mechanisms for neutrinos and ALPs in magnetars.  For neutrino emission, the dominant production mode is the modified URCA mechanism while for ALP emission, the dominant production mode is nucleon-nucleon bremsstrahlung.

\subsection{Neutrino Emissivity}

Several neutrino production mechanisms, like cyclotron emission of neutrino pairs by electrons or neutrino bremsstrahlung in the Coulomb field of ions, can lead to magnetar cooling.  However, the dominant neutrino cooling mode is the modified URCA process \cite{Friman:1978zq,Beloborodov:2016mmx},
\be\label{EqnURCA}
\dot{q}_\nu=(7\times10^{20}\,\text{erg}\cdot\text{s}^{-1}\cdot\text{cm}^{-3})\left(\frac{\rho}{\rho_0}\right)^{2/3}R_M\left(\frac{T}{10^9\,\text{K}}\right)^8,
\ee
where $\rho$ is the magnetar density, $\rho_0=2.8\times10^{14}\,\text{g}\cdot\text{cm}^{-3}$ is the nuclear saturation density, and $R_M\leq1$ is a suppression factor that appears with the onset of proton and/or neutron superfluidity.  Indeed, if superfluidity is achieved, the dominant cooling mechanism becomes Cooper pair cooling.  Although in magnetars superfluidity is not expected for protons, it is theoretically possible for neutrons when core temperatures reach $T\gtrsim10^8\,\text{K}$.  The exact critical temperature for neutron superfluidity is however not known and this issue is thus not yet settled.  To stay conservative, we therefore focus only on the not controversial and well-understood modified URCA process \eqref{EqnURCA}.

It is important to notice from \eqref{EqnURCA} that the neutrino emissivity dependence on the magnetar core temperature $T$ is quite strong.  In fact, the modified URCA process has the strongest temperature dependence of all the neutrino production mechanisms mentioned above.  Hence, within our assumptions, a small variation in the core temperature leads to a large variation in the neutrino emissivity.  Finally, for numerical purposes, we assume that $\rho=\rho_0$ and $R_M=1$ in the computation of the ALP-to-photon luminosity.

\subsection{Normalized ALP Spectrum}

There are again several ALP production mechanisms in neutron stars.  Some examples are analogs of neutrino production mechanisms like cyclotron emission of ALPs by electrons or ALP bremsstrahlung in the Coulomb field of ions.  The dominant production mechanism for ALPs is however nucleon-nucleon bremsstrahlung emission in the degenerate limit \cite{Raffelt:1996wa,Iwamoto:1984ir}, which has a normalized ALP spectrum given by
\be\label{EqnBrem}
\frac{dN_a}{d\omega}=\frac{x^2(x^2+4\pi^2)e^{-x}}{8(\pi^2\zeta_3+3\zeta_5)(1-e^{-x})},
\ee
where $x=\omega/k_BT$.  The normalized ALP spectrum \eqref{EqnBrem} satisfies $\int_0^\infty d\omega\,dN_a/d\omega=1$ and does not depend on the ALP-nucleon coupling constant $g_{aN}$.  Therefore, the normalized ALP spectrum is useful since we do not need to specify the exact model leading to ALP emission from nucleon-nucleon bremsstrahlung.  Indeed, with the cooling argument demanding that ALP luminosity does not overtake neutrino luminosity, we can stay quite general with respect to the exact type of ALP we are constraining.  For example, the ALP-nucleon coupling constant, which is model-dependent, is not needed.

However, to ensure that the cooling argument presented here, where ALP emission from nucleon-nucleon bremsstrahlung does not overcome neutrino emission from the modified URCA process, is plausible, it is nevertheless necessary to compare ALP emissivity with neutrino emissivitiy \eqref{EqnURCA}.  Although this is model-dependent, there are bounds on $g_{aN}$ and our constraint would not be as interesting if the cooling argument was already excluded by these bounds.  Following \cite{Raffelt:1996wa,Iwamoto:1984ir}, the ALP emissivity is
\be\label{EqnqBrem}
\dot{q}_a=(1.3\times10^{19}\,\text{erg}\cdot\text{s}^{-1}\cdot\text{cm}^{-3})\left(\frac{g_{aN}}{10^{-10}\,\text{GeV}^{-1}}\right)^2\left(\frac{\rho}{\rho_0}\right)^{1/3}\left(\frac{T}{10^9\,\text{K}}\right)^6,
\ee
and thus the ALP emissivity \eqref{EqnqBrem} is larger than (smaller than) [equal to] the neutrino emissivity \eqref{EqnURCA} if $g_{aN}>g_{aN}^*$ ($g_{aN}<g_{aN}^*$) [$g_{aN}=g_{aN}^*$] where $g_{aN}^*$ is given by
\be\label{EqngaN}
\left(\frac{g_{aN}^*}{10^{-10}\,\text{GeV}^{-1}}\right)=7.3\left(\frac{\rho}{\rho_0}\right)^{1/6}\sqrt{R_M}\left(\frac{T}{10^9\,\text{K}}\right).
\ee
Therefore, magnetar cooling by ALP emission is subdominant to cooling by neutrino emission if the ALP-nucleon coupling constant $g_{aN}\leq g_{aN}^*$ \eqref{EqngaN}.  For $\rho=\rho_0$, $R_M=1$ and $0.6\leq(T/10^9\,\text{K})\leq3.0$, the necessary value for $g_{aN}$ corresponds to the bound of \cite{Graham:2015ouw}, validating our model-independent approach.	


\end{document}